\def\ba{\begin{eqnarray}}
\def\ea{\end{eqnarray}}
\def\lb{\label}
\def\bi{\bibitem}

\def\D{\Delta}

\def\t{\tau}

\def\g{\gamma}

\def\x{\bar{x}}

\def\rr{\parallel}
\def\pr{\perp}

\documentclass[12pt]{article}
\usepackage{t1enc}
\usepackage[utf8x]{inputenc}
\usepackage{graphicx}%

\begin{document}

\title{ Bose-Einstein correlations  and the transition  time from QCD plasma to hadrons}

\author{A.Bialas and K.Zalewski
 \\ \\ H.Niewodniczanski Institute of
Nuclear Physics\\ Polish Academy of Sciences\thanks{Address: Radzikowskiego
152, 31-342 Krakow, Poland}\\and\\ M.Smoluchowski Institute of Physics
\\Jagellonian University\thanks{Address: Reymonta 4, 30-059 Krakow, Poland;
e-mail:bialas@th.if.uj.edu.pl;}}
 \maketitle


\begin{abstract}

It is pointed out that the size of the interaction region, as determined form HBT analyses, is increased due to the transition time necessary to convert the quarks and gluons into hadrons. A rough estimate yields an increase of $R_{HBT}$ by about 15\%.

\end{abstract}

{\bf 1.} It is now well established that particle production in high energy heavy ion collisions  proceeds through the intermediate state of a quark-gluon plasma. Much effort is devoted to investigate the properties of this state and of the  transition from the plasma to the observed hadrons.  In these investigations  measurements of  quantum interference (HBT) play a particularly interesting role, being one of the very few  sources of  information about the space-time properties of the system.

The difficulty is that neither the properties of the plasma, nor the mechanism of hadron formation are sufficiently well known. One point, however, is non-controversial. This transition must take a certain time which we call the transition time.

One can invoke several origins for the transition time. The hadrons do not
interact before they are dressed. This requires a certain time  known as the
formation time. Effects of the formation time on the HBT measurements have been
recently discussed in the framework of UrQMD \cite{graf}\footnote{The
connection of the formation time to  the uncertainty principle was considered
in  \cite{alex}).}.  Even dressed constituents cannot form a hadron instantly,
however.  Stodolsky \cite{stodolsky} analyzed from this point of view the
formation of the hydrogen atom from a proton and an electron. and  adapted this
argument for the $q\bar{q}$ system. His reasoning is briefly presented as the 
second comment towards the end of the present note. Moreover, in order to
produce a white hadron from a coloured plasma, it is necessary either to
rearrange the plasma constituents into white lumps (preconfinement), or to get
rid of the unwanted colour by emitting one, or more, soft gluons. This list
could be made longer.

  The consequences of the non-vanishing transition time  are twofold. First, it increases the time after which the
  observed hadrons appear, as compared to  estimates following, e.g., from  hydrodynamics. Second, since the
  plasma created in heavy ion collisions is far from static, the formation time leads to an increase of the
  measured volume from which the hadrons emerge.

Thus it seems   interesting to investigate how the presence of  the formation time changes the  space-time structure of the system and how  it  influences the results of the HBT measurements. This is the purpose of the present note in which we shall analyze these phenomena in a simplified model.
\vspace{0.2cm}

{\bf 2.} We consider the two-dimensional problem (in the transverse plane), assuming -for simplicity-  uncorrelated particle production\footnote{Corrections following from interpaticle correlations are discussed, e.g., in \cite{bzcor}}.
Denoting the single-particle Wigner function by $W(p,x)$, the observed single- and two-identical-particle momentum distributions are given by
\ba
\Omega_0(p)=\int d^2x W(p,x); \lb{dp}
\ea
\ba
\Omega(p_1,p_2)= F(p_1)F(p_2)\left[1+C(p_1,p_2)\right]
\ea
where
\ba
C(p_1,p_2)=\frac{|H(Q,K)|^2}{F(p_1)F(p_2)}  \lb{corr}
\ea
is the correlation function and
\ba
H(Q,K)=\int d^2x e^{iQx}W(K,x);\;\;\;\; F(p)= H(Q=0,p)=\int d^2 x W(p,x)  \lb{gen}
\ea
with $K=(p_1+p_2)/2$ and $Q=p_1-p_2$.

\vspace{0.2cm}

{\bf 3.}
 Our approach is purely phenomenological. Consider a final hadron. It  is formed from quarks and gluons emerging  from a relatively small "lump" of plasma (they have to be  close enough  in space to form a hadron). Following the argument presented before, we
 assume that in the  rest frame of the lump this process takes a certain time $\tau_0$. Naturally, this time may depend on the kind of the hadron.
 In the system where the lump of plasma moves  with velocity $v$, $\tau_0$ is multiplied by the corresponding Lorentz factor: $\tau=\gamma\tau_0$.
\vspace{0.2cm}

 Consider now the distribution of plasma at freeze-out, $G[\x]d^2\x$  (e.g. the one calculated from hydro),
   in the  Lorentz frame where the longitudinal momentum of the considered piece of  plasma vanishes (LCMS frame). It gives the distribution of the fluid in the transverse plane ($\x$ denotes  the transverse position vector in the plasma\footnote{As already mentioned, all vectors are two-dimensional and lie in the transverse plane of the collision}). At  each $\x$ also the velocity of plasma $v=v(\x)$ is known.
 E.g., in the simple case of the Hubble flow  we have
\ba
v(\x)\g(\x) = \mu \x  \lb{hubble}
\ea
where $\mu$ is a constant.

It should be emphasized that   $v(\x)$ is the  velocity of  the lump of plasma located at  position $\x$ and it is not identical with the velocities of the  produced hadrons. Indeed, in the rest frame of the lump, the momenta  $\hat{p}$ of the particles, though equal zero on the average, have some  spherically symmetric distribution.  For instance, in the statistical model they follow the  Boltzmann distribution. Therefore, the actual distribution is
\ba
W[\x,p]d^2\x d^2p= G[\x] \hat{U}[\x;\hat{p}] d^2\x d^2\hat{p}/\hat{E}=G[\x] U[\x;p] d^2\x d^2p/E
\ea
where $\hat{U}$ is the distribution of hadron momenta in the rest frame of the lump, and
$U$ is the same distribution expressed in terms of the momenta in  the laboratory system, where the lump moves with  velocity $v(\x)$.  The momenta are of course related by the Lorentz transformation:
\ba
\hat{p}_\rr=\g\{p_\rr-v E\}; \;\;\;\;\hat{E}=\g\{E-vp_\rr\};\;\;\;E=\sqrt{p^2+m^2};\;\;\;\; \hat{p}_\pr=p_\pr\lb{ee}
\ea
where the subscripts $\parallel$ and $\perp$ refer to the direction of $v(\x)$.

\vspace{0.2cm}

{\bf 4.} In the rest frame of the lump, a hadron is emitted at some time $\t_0$ which of course need not  be the same for all hadrons. In the laboratory system, this time is multiplied by $\g(\x)$.  During the time $\t=\g(\x)\t_0$ the lump moves in the laboratory system from the position $\x$ to the position
 \ba
x=\x+ \D \x  =\x+ v(\x)\g(\x)\t_0.  \lb{xx}
 \ea
This formula allows to express $\x$ in terms of $x$ and $\t_0$. Consequently,
 the distribution of the emission points becomes
\ba
W[x,p]d^2x d^2p/E= G[x-\D\x] U[\x;p] \left|\frac{d^2\x}{d^2x}\right| d^2x d^2p/E  \lb{fin}
\ea
where $|{d^2\x}/{d^2x}|$ is the Jacobian of the change of variables  from $\x$ to $x$.

 To see the physical consequences of (\ref{fin}) we have to evaluate the function
$H(Q,K)$ defined in (\ref{gen}). We have
\ba
H(Q,K)=\int d^2x e^{iQx} G[x-\D\x] U[\x;K] \left|\frac{d^2\x}{d^2x}\right|/E_K
\ea
where $E_K=\sqrt{K^2+m^2}$.
Changing the integration variable from $x$ back to $\x$ we obtain
\ba
H(Q,K)=\int d^2\x e^{iQ[\x+\D \x]} G[\x] U[\x;K] /E_K.
\ea
One sees from this formula that the dependence of the result on $\t_0$ comes only from the term $\D\x=v(\x)\g(\x)\t_0$, present in the exponent and multiplied by $Q$. From this observation one sees immediately that there is no  effect of the formation time  when there is no transverse flow, i.e. for a static plasma. One also sees  that the formation time does not change
the single particle momentum distributions.

\vspace{0.2cm}

{\bf 5.}  A particularly simple situation is obtained in the case of a radial  transverse flow, i.e. when  the velocity $v(\x)$ points in the same direction as $\x$ (this is the case in most hydrodynamical models\footnote{See, however, \cite{hama}}). Then we can write
\ba
v(\x)=|v(\x)|\frac{\x}{ |\x|}
\ea
and thus
\ba
H(Q,K)=\int d\x e^{iQ_{eff}\x} G[\x] U[\x;K] /E_K,  \lb{eff}
\ea
where
\ba
Q_{eff}=Q\left[1+\frac{|v(\x)|}{|\x|} \g(\x) \t_0\right].
\ea
This observation implies that the HBT radii obtained from (\ref{eff}) differ from  those obtained from the hydrodynamical calculations by  the factor $1+[|v(\x)|/|\x|] \g(\x) \t_0]$
averaged over the plasma distribution.

In case of the simple Hubble flow (\ref{hubble}) we obtain
\ba
Q_{eff} = Q[1+\mu \t_0]  \lb{hubtau}
\ea
and thus we conclude that in this case the observed $HBT$ radii should be larger than those evaluated from hydrodynamics by the factor $[1+\mu \t_0]$. Note that this result is independent of the distribution of plasma in the transverse plane.

To obtain an estimate of the size of the enhancement factor, one needs the value of the parameter $\mu$, responsible for the transverse flow. It seems reasonable to take  $\mu\approx 0.1/$fm, giving $v\g =1$ at $|\x| = R= 10$ fm,  in   agreement with the arguments of \cite{ww,lisa}. Then, taking  $\t_0=1/m_\pi$, one finds a correction of about 15\%.

To estimate the sensitivity of this result on the form of the transverse flow
we also considered the relation suggested in \cite{kisiel} \ba |v(\x)|\g(\x) =
\sinh(\mu_1 \x). \ea The results were evaluated numerically, using \ba
G(\x)=e^{-\x^2/R^2};\;\;\;\; U(\x,p)= e^{-\hat{E}/T}= e^{-\g(\x)[E-|v(\x)|
p_\rr]/T} \ea with $R=10$ fm, $T=160$ MeV.  The parameter   $\mu_1$ was taken
$0.08/$fm, adjusted to obtain   the same radii as in case of Hubble flow for
$\t_0=0$ and $\mu=0.1$. This exercise showed that the effects of the transition
time on $R_{HBT}^{side}$ are similar to those due to the Hubble flow (the
result (\ref{hubtau}) is not changed by more than a few percent). The increase
of  $R_{HBT}^{out}$ is similar, though somewhat larger. We thus conclude that,
within an error of a few percent, the effect of the formation time is not
sensitive to the  form of the transverse flow. Furthermore, this exercise shows
that, at least for the transition temperature in the region of 160 MeV, it is
also not substantially affected by the  correlation between $v(\x)$ and $K$,
induced by the the function $U[\x,K]$ and by the dependence of $\g$ and $v$ on
$\overline{x}$ (see   (\ref{ee})).

\vspace{0.2cm}

{\bf 6.}  In conclusion, using a simplified two-dimensional model,  we have investigated   the effects of the formation time of hadrons on  the results of the HBT measurements. It was shown that  the HBT radius measured from $\pi\pi$  correlations   is larger, at least  by about 15\%,  than that resulting  from the hydrodynamical evolution.

Some comments are in order.

(i) Our argument is based on a  simplified, two-dimensional model.  Therefore it applies, strictly speaking, only to the case where the longitudinal momenta of both particles are identical (and thus both  vanish in the LCMS). The corrections related to the longitudinal motion  tend to decrease the effect of the transition  time. They are not very important, however, if the momentum difference is kept below the pion mass.

(ii) The transition time we are discussing in this paper should not be identified with  the hadron formation time  estimated traditionally to be  $\sim 1/m$. First, as suggested in the introduction, there may be other processes contributing  to it. Second, it should be remembered that $\tau_0=1/m$  is only the smallest possible value consistent with the uncertainty principle. It was observed in \cite{stodolsky} that one can construct a classical  argument which points out to a hadron formation time  which is larger by a significant factor. Consider a particle as a bound state of two (constituent) quarks. It seems reasonable to accept that this state  can be considered as "formed"  when the constituents make at least one turn around each other. Then  one finds $\t_0 \approx 2\pi r /v$ where $r$ is the radius of the particle and $v$ is the velocity on the orbit. Taking $r\approx 1$ fm, and putting for $v$ its upper limit $v= 1$, one obtains $\t_0 \approx 2\pi$ fm $ \approx 4.5/ m_\pi$.
Note that this  argument suggests  that the actual value of the formation time may be related to the {\it size} of the produced hadron rather than to its mass.

(iii) The purpose of this note is only to point out the possible   effects of the formation time but not their precise estimate. A similar but technically  much more  detailed analysis of the role of  the time delay  in hadron production  can be found in \cite{kisiel}. Although the discussion in this paper refers to the production of the long-living resonances, their results may be  useful also in the quantitative studies of the  formation time effects.

(v) The transition time we are discussing here is  the time needed to form well defined hadrons from the quark-gluon plasma.
 Collisions between the produced hadrons, which may further influence the measurements of the HBT radii, are of course possible and even expected \cite{hg}. But this problem goes beyond the scope of the present note.

\vspace{0.2cm}

\textbf{Acknowledgements} We would like to thank W. Florkowski, G.Graef, P. Huovinen and S.Pratt    for helpful discussions. This work was supported in part by the grant N N202 123437 of the Polish Ministry of Science and Higher Education (2009 - 2012).

\end{document}